\documentclass[aps,twocolumn,amsmath,amssymb,floatfix,showpacs,
superscriptaddress]{revtex4}

\usepackage[dvips]{graphics}
\usepackage{color}
\definecolor{gold}{rgb}{0.85,0.66,0}
\definecolor{dblue}{rgb}{0,0,0.6}
\definecolor{dred}{rgb}{0.6,0,0}

\begin{document}

\title{\textcolor{dblue}{Magnetic Response in Mesoscopic Hubbard Rings: 
A Mean Field Study}}

\author{Santanu K. Maiti}

\email{santanu.maiti@saha.ac.in}

\affiliation{Theoretical Condensed Matter Physics Division, Saha
Institute of Nuclear Physics, Sector-I, Block-AF, Bidhannagar,
Kolkata-700 064, India}

\affiliation{Department of Physics, Narasinha Dutt College, 129
Belilious Road, Howrah-711 101, India}

\begin{abstract}
The present work proposes an idea to remove the long standing controversy
between the calculated and measured current amplitudes carried by a small
conducting ring upon the application of an Aharonov-Bohm (AB) flux $\phi$.
Within a mean field Hartree-Fock (HF) approximation we numerically 
calculate persistent current, Drude weight, low-field magnetic 
susceptibility and related issues. Our analysis may be inspiring for
studying magnetic response in nano-scale loop geometries.
\end{abstract}


\maketitle

\section{Introduction}

In mesoscopic range phase coherence of electronic states is of fundamental 
importance and the existence of dissipationless current in a mesoscopic 
conducting ring threaded by an Aharonov-Bohm (AB) flux $\phi$ is a 
spectacular consequence of quantum phase coherence. The existence of 
persistent current in mesoscopic rings has been addressed several years 
ago in the pioneering work of B\"{u}ttiker, Imry and Landauer~\cite{butt}. 
Later, many excellent experiments~\cite{levy,chand,jari,deb} have been 
done in different systems to reveal the phenomenon of persistent current.
Though in literature many theoretical~\cite{cheu1,cheu2,peeters1,peeters2,
peeters3,mont,alts,von,schm,ambe,bouz,giam,yu,san1,san2,san3,san4} as well 
as experimental papers~\cite{levy,chand,jari,deb} on persistent current 
are available, yet 
lot of controversies are still present between the theory and experiment, 
and the complete knowledge about it in this scale is not very well 
established even today. The unexpectedly large amplitudes of measured 
currents provide to the most important controversy. It has been proposed 
that the electron-electron (e-e) interactions contribute significantly 
to the average currents. An explanation based on the perturbative 
calculation in presence of interaction and disorder has been proposed 
and it seems to give a quantitative estimate closer to the experimental 
results, but still it is less than the measured currents by an order of 
magnitude, and the interaction parameter used in the theory is not well 
understood physically. Though an attempt has been made to explain the 
enhancement of current amplitude by some theoretical arguments but the 
sign of low-field currents cannot be predicted precisely and it is an 
important controversial issue between theoretical and experimental results. 

Motivated with these open challenges in the present paper we address 
magnetic response in mesoscopic Hubbard rings threaded by AB flux $\phi$. 
We try to propose an idea to remove the unexpected discrepancy between 
the calculated and measured current amplitudes by incorporating the 
effect of second-neighbor hopping (SNH) in addition to the traditional 
nearest-neighbor hopping (NNH) integral in the tight-binding Hamiltonian. 
Using a generalized Hartree-Fock (HF) 
approximation~\cite{kato,kam,sil}, we numerically compute persistent 
current ($I$), Drude weight ($D$) and low-field magnetic susceptibility 
($\chi$) as functions of AB flux $\phi$, total number of electrons $N_e$ 
and system size $N$. With this (HF) approach one can 
study magnetic response in a much larger system since here a many-body
Hamiltonian is decoupled into two effective one-body Hamiltonians. One 
is associated with up spin electrons and other is related to down spin 
electrons. But the point is that, the results calculated using generalized 
HF mean-field theory may deviate from exact results with the reduction of 
\begin{figure}[ht]
{\centering \resizebox*{4.3cm}{2.5cm}{\includegraphics{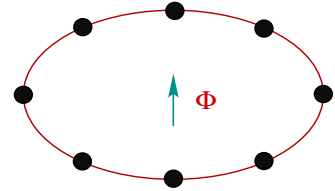}}\par}
\caption{(Color online). Schematic view of a $1$D mesoscopic ring
penetrated by a magnetic flux $\phi$. The filled black circles
correspond to the positions of the atomic sites.}
\label{ring}
\end{figure}
dimensionality. So we should take care about the mean-field calculation, 
specially, in one-dimensional systems. To trust our predictions, in the 
present work also we make a comparative study between the results obtained 
from mean-field theory and exactly diagonalizing the full many-body 
Hamiltonian. The later approach where a complete many-body Hamiltonian
is diagonalized to get energy eigenvalues is not suitable to study magnetic
response in larger systems since the size of the matrices increases very
sharply with the total number of up and down spin electrons.
Our results can be utilized to explore magnetic 
response in any interacting mesoscopic system. 

We organize the paper as follows. Following a brief introduction (Section
I), in Section II we describe the geometric model and generalized
Hartree-Fock theory to study magnetic response in the model quantum system.
Section III contains the numerical results, and finally, summary of our
work will be available in Section IV.

\section{Model and theoretical formulation}

We start by referring to Fig.~\ref{ring}, where a normal metal ring 
is threaded by a magnetic flux $\phi$. 
To describe the system we use a tight-binding framework. For a $N$-site
ring, penetrated by a magnetic flux $\phi$ (measured in unit of the 
elementary flux quantum $\phi_0=ch/e$), the tight-binding Hamiltonian
in Wannier basis looks in the form,
\begin{eqnarray}
H_R & = &\sum_{i,\sigma}\epsilon_{i\sigma} c_{i\sigma}^{\dagger} c_{i\sigma}
+\sum_{ij,\sigma} t \left[e^{i\theta} c_{i\sigma}^{\dagger} 
c_{j\sigma} + h.c. \right] + \nonumber \\
&  & \sum_{ik,\sigma} t_1 \left[e^{i\theta_1} c_{i\sigma}^{\dagger} 
c_{k\sigma} + h.c. \right] + \sum_i U c_{i\uparrow}^{\dagger}c_{i\uparrow} 
c_{i\downarrow}^{\dagger} c_{i\downarrow}
\label{equ1}
\end{eqnarray}
where, $\epsilon_{i\sigma}$ is the on-site energy of an electron at 
the site $i$ of spin $\sigma$ ($\uparrow,\downarrow$). The variable $t$ 
corresponds to the nearest-neighbor ($j=i\pm1$) hopping strength, while
$t_1$ gives the second-neighbor ($k=i\pm2$) hopping integral. 
$\theta=2\pi\phi/N$ and $\theta_1=4\pi\phi/N$ are the phase factors
associated with the hopping of an electron from one site to its neighboring
site and next-neighboring site, respectively. $c_{i\sigma}^{\dagger}$ and 
$c_{i\sigma}$ are the creation and annihilation operators, respectively, 
of an electron at the site $i$ with spin $\sigma$. $U$ is the strength of 
on-site Hubbard interaction. 

\subsection{Decoupling of the interacting Hamiltonian}

In order to determine the energy eigenvalues of the interacting model 
quantum system described by the tight-binding Hamiltonian given in
Eq.~\ref{equ1}, first we decouple the interacting Hamiltonian using 
generalized Hartree-Fock approach, the so-called mean field approximation.
In this approach, the full Hamiltonian is completely decoupled into two
parts. One is associated with the up-spin electrons, while the other is
related to the down-spin electrons with their modified site energies.
For up and down spin Hamiltonians, the modified site energies are 
expressed in the form,
$\epsilon_{i\uparrow}^{\prime}=\epsilon_{i\uparrow} + U \langle 
n_{i\downarrow} \rangle$ and
$\epsilon_{i\downarrow}^{\prime}=\epsilon_{i\downarrow} + U \langle 
n_{i\uparrow} \rangle$,
where $n_{i\sigma}=c_{i\sigma}^{\dagger} c_{i\sigma}$ is the number
operator. With these site energies, the full Hamiltonian (Eq.~\ref{equ1})
can be written in the decoupled form as,
\begin{eqnarray}
H_R &=&\sum_i \epsilon_{i\uparrow}^{\prime} n_{i\uparrow} + \sum_{ij} t 
\left[e^{i\theta} c_{i\uparrow}^{\dagger} c_{j\uparrow} + e^{-i\theta} 
c_{j\uparrow}^{\dagger} c_{i\uparrow}\right] \nonumber \\
& + & \sum_{ik} t_1 \left[e^{i\theta_1} c_{i\uparrow}^{\dagger} c_{k\uparrow}
+ e^{-i\theta_1} c_{k\uparrow}^{\dagger} c_{i\uparrow}\right] \nonumber \\
& + & \sum_i \epsilon_{i\downarrow}^{\prime} n_{i\downarrow} + \sum_{ij} 
t \left[e^{i\theta} c_{i\downarrow}^{\dagger} c_{j\downarrow}
+ e^{-i\theta} c_{j\downarrow}^{\dagger} c_{i\downarrow}\right] \nonumber \\
& + & \sum_{ik} t_1 \left[e^{i\theta_1} c_{i\downarrow}^{\dagger} 
c_{k\downarrow} + e^{-i\theta_1} c_{k\downarrow}^{\dagger} c_{i\downarrow}
\right] \nonumber \\
& - & \sum_i U \langle n_{i\uparrow} \rangle \langle n_{i\downarrow} 
\rangle \nonumber \\
&=& H_{\uparrow}+H_{\downarrow}-\sum_i U \langle n_{i\uparrow} 
\rangle \langle n_{i\downarrow} \rangle
\label{equ4} 
\end{eqnarray}
where, $H_{\uparrow}$ and $H_{\downarrow}$ correspond to the effective
tight-binding Hamiltonians for the up and down spin electrons, respectively.
The last term is a constant term which provides an energy shift in the 
total energy. 

\subsection{Self consistent procedure}

With these decoupled Hamiltonians ($H_{\uparrow}$ and $H_{\downarrow}$) 
of up and down spin electrons, now we start our self consistent procedure 
considering initial guess values of $\langle n_{i\uparrow} \rangle$ and 
$\langle n_{i\downarrow} \rangle$. For these initial set of values of
$\langle n_{i\uparrow} \rangle$ and $\langle n_{i\downarrow} \rangle$, 
we numerically diagonalize the up and down spin Hamiltonians. Then we 
calculate a new set of values of $\langle n_{i\uparrow} \rangle$ and 
$\langle n_{i\downarrow} \rangle$. These steps are repeated until a self
consistent solution is achieved.

\subsection{Calculation of ground state energy}

Using the self consistent solution, the ground state energy 
$E_0$ for a particular filling at absolute zero temperature ($T=0$K)
can be determined by taking the sum of individual states up to Fermi 
energy ($E_F$) for both up and down spins. Thus, we can write the final
form of ground state energy as,
\begin{equation}
E_0=\sum_n E_{n\uparrow} + \sum_n E_{n\downarrow}- \sum_i U \langle
n_{i\uparrow} \rangle \langle n_{i\downarrow} \rangle
\label{equ5} 
\end{equation} 
where, the index $n$ runs for the states upto the Fermi level. 
$E_{n\uparrow}$ ($E_{n\downarrow}$) is the single particle energy 
eigenvalue for $n$-th eigenstate obtained by diagonalizing the
Hamiltonian $H_{\uparrow}$ ($H_{\downarrow}$).

\subsection{Calculation of persistent current} 

At absolute zero temperature, total persistent current of the system
is obtained from the expression,
$I(\phi)=-c~\partial E_0(\phi)/\partial \phi$
where, $E_0(\phi)$ is the ground state energy.

\subsection{Calculation of Drude weight} 

The Drude weight for the ring can be calculated through the relation,
\begin{equation}
D=\left . \frac{N}{4\pi^2} \left(\frac{\partial{^2E_0(\phi)}}
{\partial{\phi}^{2}}\right) \right|_{\phi \rightarrow 0}
\label{equ7}
\end{equation}
where, $N$ gives total number of atomic sites in the ring. Kohn~\cite{kohn} 
has shown that for an insulating system $D$ decays exponentially to zero, 
while it becomes finite for a conducting system.

\subsection{Determination of low-field magnetic susceptibility} 

The general expression of magnetic susceptibility $\chi$ at any flux 
$\phi$ is written in the form,
\begin{equation}
\chi(\phi)=\frac{N^3}{16\pi^2}\left(\frac{\partial I(\phi)}{\partial \phi}
\right).
\label{equ8}
\end{equation}
Evaluating the sign of $\chi(\phi)$ we can able to predict whether the
current is paramagnetic or diamagnetic in nature. Here we will determine
$\chi(\phi)$ only in the limit $\phi \rightarrow 0$, since we are
interested to know the magnetic response in the low-field limit.

In the present work we perform all the essential features of magnetic
response at absolute zero temperature and use the units where $c=h=e=1$.
Throughout our numerical work we set the nearest-neighbor hopping strength 
$t=-1$ and second-neighbor hopping strength $t_1=-0.7$. Energy scale is 
measured in unit of $t$.

\section{Numerical results and discussion}

\subsection{Perfect Hubbard Rings Described with NNH Integral}

For perfect rings we choose $\epsilon_{i \uparrow}=
\epsilon_{i \downarrow}=0$ for all $i$ and since here we consider the 
rings with only NNH integral, the second-neighbor hopping strength $t_1$
is fixed at zero.

\subsubsection{Energy-flux characteristics}

\begin{figure}[ht]
{\centering \resizebox*{7.75cm}{8cm}{\includegraphics{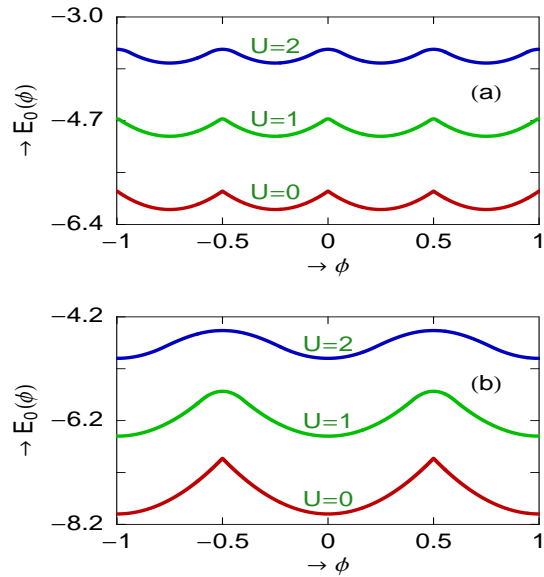}}\par}
\caption{(Color online). Ground state energy levels as a function of flux 
$\phi$ for some typical mesoscopic rings in half-filled case. The red, 
green and blue curves correspond to $U=0$, $1$ and $2$, respectively. 
(a) $N=5$ and (b) $N=6$.}
\label{ringenergy}
\end{figure}

To explain the relevant features of magnetic response we begin with
the energy-flux characteristics. As illustrative examples, in 
Fig.~\ref{ringenergy} we plot the ground state energy levels as a 
function of magnetic flux $\phi$ for some typical mesoscopic rings in 
the half-filled case, where (a) and (b) correspond to $N=5$ and $6$, 
respectively. The red curves represent the energy levels for the 
non-interacting ($U=0$) rings, while the green and blue lines correspond 
to the energy levels for the interacting rings where the electronic 
correlation strength $U$ is
fixed to $1$ and $2$, respectively. From the spectra it is observed that
the ground state energy level shifts towards the positive energy and it
becomes much flatter with the increase of the correlation strength $U$. 
Both for the two different ring sizes ($N=5$ and $6$) the ground state
energy levels vary periodically with AB flux $\phi$, but a significant
difference is observed in their periodicities depending on the oddness
and evenness of the ring size $N$. For $N=6$ (even), the energy levels 
show conventional $\phi_0$ ($=1$, in our chosen unit system $c=e=h=1$) 
flux-quantum periodicity. On the other hand, the period becomes half i.e., 
$\phi_0/2$ for $N=5$ (odd). This $\phi_0/2$ periodicity disappears as long 
as the filling is considered away from the half-filling. At the same time, 
it also vanishes if impurities are introduced in the system, even if the 
ring is half-filled with odd $N$. Therefore, $\phi_0/2$ periodicity is a 
special feature for odd half-filled perfect rings irrespective of the 
Hubbard strength $U$, while for all other cases traditional $\phi_0$ 
periodicity is obtained. 

\begin{figure}[ht]
{\centering \resizebox*{7.75cm}{8cm}{\includegraphics{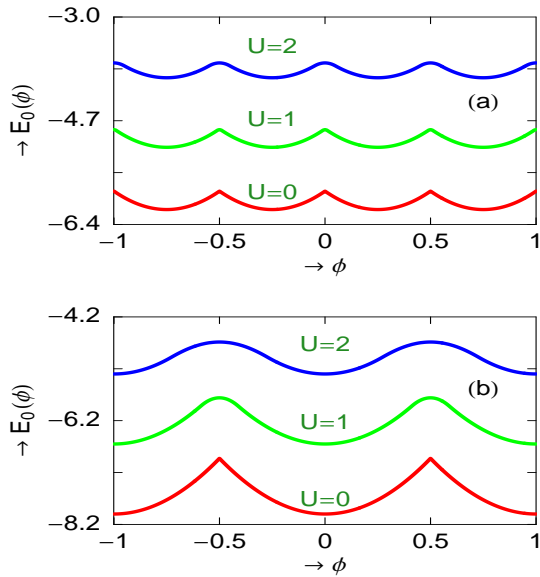}}\par}
\caption{(Color online). Ground state energy levels as a function of flux 
$\phi$ for some typical mesoscopic rings in half-filled case, where 
eigenenergies are determined through exact diagonalization of the full
many-body Hamiltonian. The red, green and blue curves correspond to $U=0$, 
$1$ and $2$, respectively. (a) $N=5$ and (b) $N=6$.}
\label{exactenergy}
\end{figure}

To judge the accuracy of the mean-field calculations 
in our ring geometry, in Fig.~\ref{exactenergy} we show the variation of 
lowest energy levels where the eigenenergies are determined through exact 
diagonalization of the full many-body Hamiltonian for the identical rings 
as given in Fig.~\ref{ringenergy}, considering the same parameter values.
Comparing the results presented in Figs.~\ref{ringenergy} and 
\ref{exactenergy}, we see that the mean-field results agree very well 
with the exact diagonalization method. Thus we can safely use mean-field
approach to study magnetic response in our geometry.

\subsubsection{Current-flux characteristics}

\begin{figure}[ht]
{\centering \resizebox*{7.75cm}{8cm}{\includegraphics{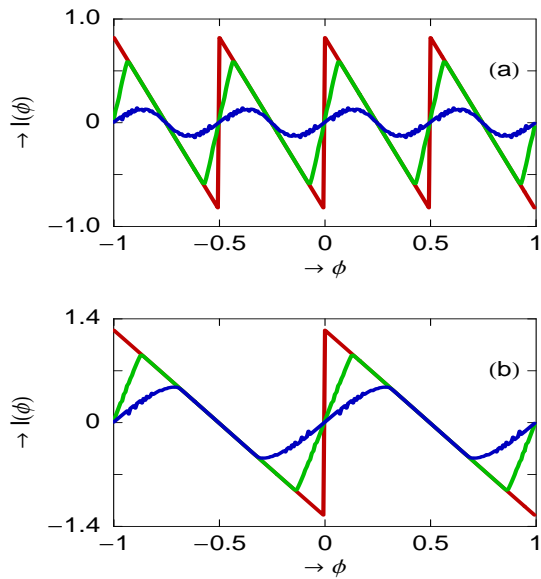}}\par}
\caption{(Color online). Persistent current as a function of flux $\phi$ 
for single-channel mesoscopic rings in half-filled case. (a) $N=15$. The 
red, green and blue curves correspond to $U=0$, $1.5$ and $2$, respectively. 
(b) $N=20$. The red, green and blue curves correspond to $U=0$, $1$ and 
$1.5$, respectively.}
\label{ringcurr}
\end{figure}

Following the above energy-flux characteristics now we describe the
behavior of persistent current in mesoscopic Hubbard rings. As 
representative examples, in Fig.~\ref{ringcurr} we display the variation
of persistent currents as a function of flux $\phi$ for some typical 
single-channel mesoscopic rings in the half-filled case, where (a) and 
(b) correspond to $N=15$ and $20$, respectively. The red, green and blue
curves in Fig.~\ref{ringcurr}(a) correspond to the currents for $U=0$,
$1.5$ and $2$, respectively, while these curves in Fig.~\ref{ringcurr}(b)
represent the currents for $U=0$, $1$ and $1.5$, respectively. In the
absence of any e-e interaction ($U=0$), persistent current shows
saw-tooth like nature as a function of flux $\phi$ with sharp transitions
at $n\phi_0/2$ (red line of Fig.~\ref{ringcurr}(a)) or $n\phi_0$ (red
line of Fig.~\ref{ringcurr}(b)), where $n$ being an integer, depending on
whether $N$ is odd or even. The saw-tooth like behavior disappears as 
long as the electronic correlation is introduced into the system. This 
is clearly observed from the green and blue curves of Fig.~\ref{ringcurr}.
Additionally, in the presence of $U$, the current amplitude gets
suppressed compared to the current amplitude in the non-interacting case,
and it decreases gradually with increasing $U$. This provides the lowering
of electron mobility with the rise of $U$ and the reason behind this can
be much better understood from our forthcoming discussion. Both for two
different rings with sizes $N=15$ (odd) and $20$ (even), persistent 
currents vary periodically with AB flux $\phi$ showing different 
periodicities, following the energy-flux characteristics. For $N=15$,
current shows $\phi_0/2$ flux-quantum periodicity, while for the other 
case ($N=20$), current exhibits $\phi_0$ flux-quantum periodicity.

\subsubsection{Variation of electronic mobility: Drude weight}

To reveal the conducting properties of Hubbard rings, we study the 
variation of Drude weight $D$ for these systems. Drude weight can be
calculated by using Eq.~\ref{equ7}. Finite value of $D$ predicts the 
metallic phase, while for the insulating phase it drops exponentially
to zero~\cite{kohn}.

\begin{figure}[ht]
{\centering \resizebox*{7.75cm}{8cm}{\includegraphics{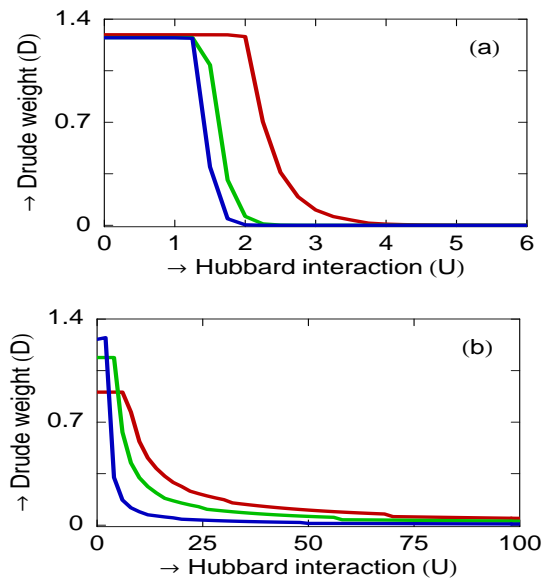}}\par}
\caption{(Color online). Drude weight as a function of Hubbard interaction
strength $U$ for single-channel mesoscopic rings. (a) Half-filled case.
The red, green and blue curves correspond to $N=10$, $30$ and $50$, 
respectively. (b) Non-half-filled case with $N=20$. The red, green and 
blue curves correspond to $N_e=10$, $14$ and $18$, respectively.}
\label{ringdrude}
\end{figure}

\begin{figure}[ht]
{\centering \resizebox*{7.75cm}{8cm}{\includegraphics{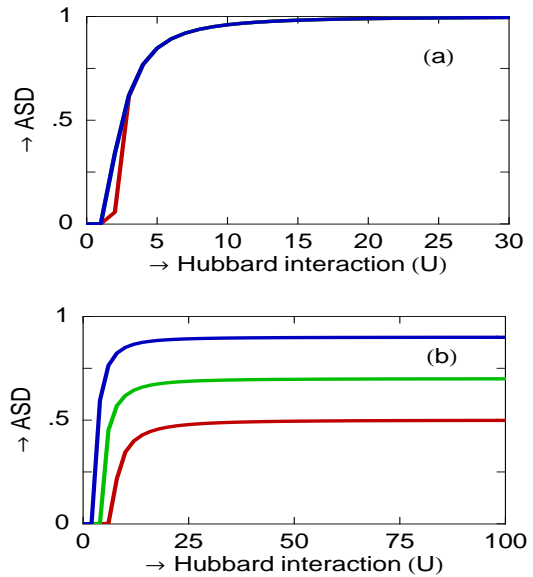}}\par}
\caption{(Color online). Average spin density (ASD) as a function of 
Hubbard interaction strength $U$ for single-channel mesoscopic rings. 
(a) Half-filled case. The red, green and blue curves correspond to $N=10$, 
$30$ and $50$, respectively. (b) Non-half-filled case with $N=20$. The red, 
green and blue curves correspond to $N_e=10$, $14$ and $18$, respectively.}
\label{ringspinden}
\end{figure}

As illustrative examples, in Fig.~\ref{ringdrude} we show the variation
of Drude weight $D$ as a function of electronic correlation strength $U$
for some typical single-channel Hubbard rings. In Fig.~\ref{ringdrude}(a)
the results are shown for three different half-filled rings, where the
red, green and blue lines correspond to the rings with $N=10$, $30$ and 
$50$, respectively. From the curves it is evident that for smaller 
values of $U$, the half-filled rings show finite value of $D$ which 
reveals that they are in the metallic phase. On the other hand, $D$
drops sharply to zero when $U$ becomes high. Thus the rings become
insulating when $U$ is quite large. The results for the non-half filled
case are shown in Fig.~\ref{ringdrude}(b), where we fix the ring size 
$N=20$ and vary the electron filling. The red, green and blue curves
represent $N_e=10$, $14$ and $18$, respectively, where $N_e$ gives the
total number of electrons in the ring. For these three choices of $N_e$,
the ring is always less than half-filled (since $N_e<N$) and the ring 
is in the conducting phase irrespective of the correlation strength $U$.
Now we try to justify the dependence of the Hubbard strength $U$ on the
electronic mobility for these different fillings. To understand the 
effect of $U$ on electron mobility here we measure a quantity called
`average spin density' (ASD) which is defined by the factor $\sum_i 
|(n_{i\uparrow}-n_{i\downarrow})|/N$. The integer $i$ is the site index
and it runs from $1$ to $N$. By calculating ASD we can estimate the 
occupation probability of electrons in the ring and it supports us
to explain whether the ring lies in the metallic phase or in the 
insulating one. For the rings those are below half-filled, ASD is always 
less than unity irrespective of the value of $U$ as shown by the curves
in Fig.~\ref{ringspinden}(b). It reveals that for these systems, ground
state always supports an empty site and electron can move along the
ring avoiding double occupancy of two different spin electrons at any 
site $i$ in the presence of e-e correlation which provides the metallic 
phase ($D>0$). For a fixed ring size and a particular strength of $U$, 
the ASD increases as the filling is increased towards half-filling which
is noticed by comparing the three different curves in 
Fig.~\ref{ringspinden}(b). On the other hand, in the half-filled rings, 
ASD is less than unity for small value of $U$, while it reaches to unity 
when $U$ is large. This behavior is clearly shown by the curves given in 
Fig.~\ref{ringspinden}(a), where the red, green and blue lines correspond 
to ASDs for the half-filled rings with $N=10$, $30$ and $50$, respectively. 
Thus, for low $U$ there is some finite probability of getting two opposite 
spin electrons in a same site which allows electrons to move along the ring
and the metallic phase is obtained. But for large $U$, ASD reaches to
unity which means that each site is singly occupied either by an up or
down spin electron with probability $1$. In this case ground state
does not support any empty site and due to strong repulsive e-e 
correlation one electron sitting in a site does not allow to come other
electron with opposite spin from the neighboring site which provides the
insulating phase ($D=0$). The situation is somewhat analogous to Mott 
localization in one-dimensional infinite lattices.
In perfect Hubbard rings the conducting nature has been
studied exactly quite a long ago using the ansatz of Bethe by Shastry
and Sutherland~\cite{shastry}. They have calculated charge stiffness 
constant ($D_c$) and have predicted that $D_c$ goes to zero as the system 
approaches towards half-filling for any non-zero value of $U$. Our 
numerical results clearly justify their findings.

\subsubsection{Low-field magnetic susceptibility}

\begin{figure}[ht]
{\centering \resizebox*{7.75cm}{8cm}{\includegraphics{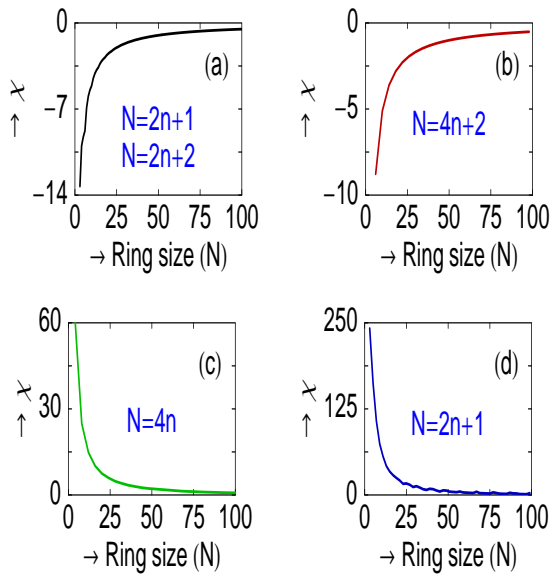}}\par}
\caption{(Color online). Low-field magnetic susceptibility as a function of 
system size $N$ for single-channel mesoscopic rings in half-filled case. 
(a) $U=0$. $N$ is an odd ($2n+1$) or an even ($2n+2$) number, where $n$ is
an integer. (b) $U=1$. $N$ is an even number obeying the relation $N=4n+2$. 
(c) $U=1$. $N$ is an even number satisfying the relation $N=4n$. (d) $U=1$.
$N$ is an odd number following the relation $N=2n+1$.}
\label{ringsuscep}
\end{figure}

Now, we discuss the variation of low-field magnetic susceptibility which 
can be calculated from Eq.~\ref{equ8} by setting $\phi \rightarrow 0$. 
With the help of this parameter we can justify
whether the current is paramagnetic ($+$ve slope) or diamagnetic ($-$ve
slope) in nature. For our illustrative purposes, in Fig.~\ref{ringsuscep}
we show the variation of low-field magnetic susceptibility with system 
size $N$ for some typical single-channel mesoscopic rings in the 
half-filled case. Figure~\ref{ringsuscep}(a) correspond to the variation
of low-field magnetic susceptibility for the non-interacting ($U=0$) rings,
where the ring size can by anything i.e., either odd, following the
relation $N=2n+1$ ($n$ is an integer), or even, obeying the expression
$N=2n+2$. It is observed that both for odd and even $N$, low-field current
exhibits diamagnetic nature. The behavior of the low-field currents changes
significantly when the e-e interaction is taken into account. Depending
on the ring size $N$, the sign becomes $+$ve and $-$ve as shown by the
curves given in Figs.~\ref{ringsuscep}(b)-(d). For the interacting rings
where the relation $N=4n+2$ is satisfied, the low-field current becomes
diamagnetic (Fig.~\ref{ringsuscep}(b)). The sign becomes paramagnetic when 
$N=4n$ (Fig.~\ref{ringsuscep}(c)) and $N=2n+1$ (Fig.~\ref{ringsuscep}(d)).
Thus, in brief, we say that for non-interacting half-filled rings low-field 
current exhibits diamagnetic response irrespective of $N$ i.e., whether
$N$ is odd or even. For the interacting half-filled rings with odd $N$,
low-field current provides only the paramagnetic behavior, while for
even $N$, depending on the particular value of $N$, the response becomes 
either diamagnetic or paramagnetic. These natures of low-field currents
change for the cases of other electron fillings. Hence, it can be
emphasized that the behavior of the low-field currents is highly sensitive
on the Hubbard correlation, electron filling, evenness and oddness of $N$, 
etc. The behavior of zero-field magnetic susceptibility
in Hubbard rings has been studied extensively quite a long back using the 
Bethe ansatz by Shiba~\cite{shiba}. In this work, he has studied magnetic
susceptibility per electron as functions of electron filling and Hubbard 
correlation strength and provided several interesting results. From his
findings we can clearly justify our presented results.

\subsection{Disordered Hubbard Rings Described with NNH and SNH 
Integrals}

Finally, we explore the combined effect of electron-electron correlation
and second-neighbor hopping (SNH) integral on persistent current in 
disordered mesoscopic rings.

To get a disordered ring, we choose site energies ($\epsilon_{i\uparrow}$
and $\epsilon_{i\downarrow}$) randomly from a ``Box" distribution function 
of width $W$. As the site energies are chosen randomly it is needed to 
consider the average over a large number of disordered configurations (from 
the stand point of statistical average). Here, we determine the currents by 
taking the average over $50$ random disordered configuration in each case 
to achieve much accurate results.

\begin{figure}[ht]
{\centering \resizebox*{7.75cm}{8cm}{\includegraphics{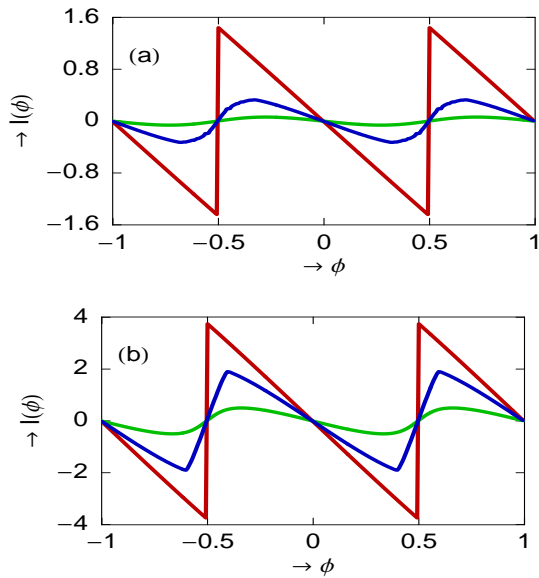}}\par}
\caption{(Color online). Persistent current as a function of flux $\phi$ 
for single-channel mesoscopic rings with $N=15$ considering $1/3$ electron
filling. (a) Rings with only NNH integral. The red line corresponds to the
ordered non-interacting ring, while the green and blue lines correspond 
to the disordered ($W=2$) rings with $U=0$ and $0.5$, respectively. (b) 
Rings with NNH and SNH integrals. The red line represents the ordered 
non-interacting ring, whereas the green and blue line correspond to the 
disordered ($W=2$) rings with $U=0$ and $1.5$, respectively.}
\label{ringcurrsnh}
\end{figure}

As illustrative examples, in Fig.~\ref{ringcurrsnh} we display the 
variation of persistent currents for some single-channel mesoscopic 
rings considering $1/3$ electron filling. In (a) the results are given
for the rings characterized by the NNH integral model. The red curve 
represents the current for the ordered ($W=0$) non-interacting ($U=0$) 
ring. It shows saw-tooth like nature with AB flux $\phi$ providing $\phi_0$ 
flux-quantum periodicity. The situation becomes completely different when 
impurities are introduced in the ring as clearly seen by the other two
colored curves. The green curve represents the current for the case only 
when impurities are considered but the effect of Hubbard interaction is 
not taken into account. It varies continuously with $\phi$ and gets much 
reduced amplitude, even an order of magnitude, compared to the perfect case.
This is due to the localization of the energy eigenstates in the presence
of impurity, which is the so-called Anderson localization. Hence, a large
difference exists between the current amplitudes of an ordered and 
disordered non-interacting rings and it was the main controversial issue 
among the theoretical and experimental predictions. Experimental results
suggest that the measured current amplitude is quite comparable to the
theoretically estimated current amplitude in a perfect system. To remove 
this controversy, as a first attempt, we include the effect of Hubbard
interaction in the disordered ring described by the NNH model. The result 
is shown by the blue curve where $U$ is fixed at $0.5$. It is observed 
that the current amplitude gets increased compared to the non-interacting 
disordered ring, though the increment is too small. Not only that the 
enhancement can take place only for small values of $U$, while for large 
enough $U$ the current amplitude rather decreases. This phenomenon can 
be explained as follows. For the non-interacting disordered ring the 
probability of getting two opposite spin electrons becomes higher at 
the atomic sites where the site energies are lower than the other sites 
since the electrons get pinned at the lower site energies to minimize 
the ground state energy, and this pinning of electrons becomes increased 
with the rise of impurity strength $W$. As a result the mobility of 
electrons and hence the current amplitude gets reduced with the increase 
of impurity strength $W$. Now, if we introduce electronic correlation in 
the system then it tries to depin two opposite spin electrons those are 
situated together due to the Coulomb repulsion. Therefore, the electronic 
mobility is enhanced which provides quite larger current amplitude. But, 
for large enough interaction strength, mobility of electrons gradually
decreases due to the strong repulsive interaction. Accordingly, the current 
amplitude gradually decreases with $U$. So, in short, we can say that
within the nearest-neighbor hopping (NNH) model electron-electron
interaction does not provide any significant contribution to enhance
the current amplitude, and hence the controversy regarding the current
amplitude still persists.

To overcome this controversy, finally we make an attempt by incorporating 
the effect of second-neighbor hopping (SNH) integral in addition to the 
nearest-neighbor hopping (NNH) integral. With this modification a 
significant change in current amplitude takes place which is clearly 
observed from Fig.~\ref{ringcurrsnh}(b). The red curve refers 
to the current for the perfect ($W=0$) non-interacting ($U=0$) ring
and it achieves much higher amplitude compared to the NNH model (see red 
curve of Fig.~\ref{ringcurrsnh}(a)). This additional contribution
comes from the SNH integral since it allows electrons to hop further.
The main focus of this sub-section is to interpret the combined effect
of SNH integral and Hubbard correlation on the enhancement of persistent 
current in disordered ring. To do this first we narrate the effect of 
SNH integral in disordered non-interacting ring.
The nature of the current for this particular case is shown by the green
curve of Fig.~\ref{ringcurrsnh}(b). It shows that the current amplitude
gets reduced compared to the perfect case (red line), which is expected,
but the reduction of the current amplitude is very small than the NNH
integral model (see green curve of Fig.~\ref{ringcurrsnh}(a)). This is
due the fact that the SNH integral tries to delocalize the electronic
states, and therefore, the mobility of the electrons is enriched. The
situation becomes more interesting when we include the effect of Hubbard
interaction. The behavior of the current in the presence of interaction
is plotted by the blue curve of Fig.~\ref{ringcurrsnh}(b) where we fix
$U=1.5$. Very interestingly we see that the current amplitude is enhanced
moderately and quite comparable to that of the perfect cylinder. Therefore, 
it can be predicted that the presence of SNH integral and Hubbard
interaction can provide a persistent current which may be comparable to the
measured current amplitudes. In this presentation we consider the effect of
only SNH integral in addition to the NNH model, and, illustrate how such 
a higher order hopping integral leads an important role on the enhancement 
of current amplitude in presence of Hubbard correlation for disordered
rings. Instead of considering only the SNH integral we can also take
the contributions from all possible higher order hopping integrals with
reduced hopping strengths. Since the strengths of other higher order
hopping integrals are too small, the contributions from these factors
are reasonably small and they will not provide any significant change in
the current amplitude. Finally, we can say that further studies are
needed by incorporating all these factors.

\section{Closing remarks}

To summarize, we have studied magnetic response in mesoscopic Hubbard 
rings threaded by Aharonov-Bohm flux $\phi$. Using the generalized 
Hartree-Fock (HF) approximation, we have numerically computed persistent 
current, Drude weight, low-field magnetic susceptibility and some
other related issues. Most importantly, we have tried to implement
an idea to remove the long standing problem between the experimentally 
measured and theoretically predicted current amplitudes. We believe
the present analysis is found to exhibit several interesting results 
which have so far remained unaddressed.

\vskip 0.3in
\noindent
{\bf\small ACKNOWLEDGMENTS}
\vskip 0.2in
\noindent
I acknowledge with deep sense of gratitude the illuminating comments
and suggestions I have received from Prof. Shreekantha Sil, Prof.
S. N. Karmakar, Prof. Bibhas Bhattacharyya, Srilekha Saha and Moumita
Dey during the calculations.


\begin{thebibliography}{99}

\bibitem{butt} M. B\"{u}ttiker, Y. Imry, and R. Landauer, Phys. Lett.
\textbf{96A}, 365 (1983).
\bibitem{levy} L. P. Levy, G. Dolan, J. Dunsmuir, and H. Bouchiat,
Phys. Rev. Lett. \textbf{64}, 2074 (1990).
\bibitem{chand} V. Chandrasekhar, R. A. Webb, M. J. Brady, M. B. Ketchen,
W. J. Gallagher, and A. Kleinsasser, Phys. Rev. Lett. \textbf{67},
3578 (1991).
\bibitem{jari} E. M. Q. Jariwala, P. Mohanty, M. B. Ketchen, and R. A. Webb,
Phys. Rev. Lett. \textbf{86}, 1594 (2001).
\bibitem{deb} R. Deblock, R. Bel, B. Reulet, H. Bouchiat, and D. Mailly,
Phys. Rev. Lett. \textbf{89}, 206803 (2002).
\bibitem{cheu1} H. F. Cheung, Y. Gefen, E. K. Riedel, and W. H. Shih,
Phys. Rev. B \textbf {37}, 6050 (1988).
\bibitem{cheu2} H. F. Cheung, E. K. Riedel, and Y. Gefen, Phys. Rev. Lett.
\textbf{62}, 587 (1989).
\bibitem{peeters1} L. K. Castelano, G.-Q. Hai, B. Partoens, and F. M. 
Peeters, Phys. Rev. B \textbf{78}, 195315 (2008). 
\bibitem{peeters2} M. Zarenia, M. J. Pereira, F. M. Peeters, and G. de 
Farias, Phys. Rev. B \textbf{81}, 045431 (2010). 
\bibitem{peeters3} D. Y. Vodolazov, F. M. Peeters, T. T. Hongisto, and 
K. Yu. Arutyunov, Europhys. Lett. \textbf{75}, 315 (2006). 
\bibitem{mont} G. Montambaux, H. Bouchiat, D. Sigeti, and R. Friesner,
Phys. Rev. B \textbf{42}, 7647 (1990).
\bibitem{alts} B. L. Altshuler, Y. Gefen, and Y. Imry, Phys. Rev. Lett.
\textbf{66}, 88 (1991).
\bibitem{von} F. von Oppen and E. K. Riedel, Phys. Rev. Lett.
\textbf{66}, 84 (1991).
\bibitem{schm} A. Schmid, Phys. Rev. Lett. \textbf{66}, 80 (1991).
\bibitem{ambe} V. Ambegaokar and U. Eckern, Phys. Rev. Lett.
\textbf{65}, 381 (1990).
\bibitem{bouz} G. Bouzerar, D. Poilblanc, and G. Montambaux, Phys.
Rev. B \textbf{49}, 8258 (1994).
\bibitem{giam} T. Giamarchi and B. S. Shastry, Phys. Rev. B \textbf{51},
10915 (1995).
\bibitem{yu} N. Yu and M. Fowler, Phys. Rev. B \textbf{45}, 11795 (1992).
\bibitem{san1} S. K. Maiti, J. Chowdhury, and S. N. karmakar, Phys.
Lett. A \textbf{332}, 497 (2004).
\bibitem{san2} S. K. Maiti, J. Chowdhury, and S. N. karmakar, J. Phys.:
Condens. Matter \textbf{18}, 5349 (2006).
\bibitem{san3} S. K. Maiti, Physica E \textbf{31}, 117 (2006).
\bibitem{san4} S. K. Maiti, Solid State Phenomena \textbf{155}, 87 (2009).
\bibitem{kato} H. Kato and D. Yoshioka, Phys. Rev. B \textbf{50}, 4943
(1994).
\bibitem{kam} A. Kambili, C. J. Lambert, and J. H. Jefferson, Phys.
Rev. B \textbf{60}, 7684 (1999).
\bibitem{sil} S. Gupta, S. Sil, and B. Bhattacharyya, Physica B 
\textbf{355}, 299 (2005).
\bibitem{kohn} W. Kohn, Phys. Rev. \textbf{133}, A171 (1964).
\bibitem{shastry} B. S. Shastry and B. Sutherland, Phys. Rev. Lett. 
\textbf{65}, 243 (1990).
\bibitem{shiba} H. Shiba, Phys. Rev. B \textbf{6}, 930 (1972).

\end{thebibliography}
\end{document}